\documentclass[twocolumn,aps,superscriptaddress]{revtex4-1}
\usepackage{palatino}
\usepackage[latin1]{inputenc}
\usepackage{epsf}
\usepackage{amsmath,amssymb}
\usepackage{latexsym}
\usepackage{calc}
\usepackage{color}
\usepackage{shadow}
\usepackage{epsfig}
\usepackage{braket}

\newcommand{\ben}{\begin{equation}}
\newcommand{\een}{\end{equation}}
\newcommand{\bea}{\begin{eqnarray}}
\newcommand{\eea}{\end{eqnarray}}

\def\sss{\scriptscriptstyle\rm}
\def\x{_{\sss X}}
\def\c{_{\sss C}}
\def\s{_{\sss S}}
\def\xc{_{\sss XC}}

\def\H{_{\sss H}}

\def\ext{_{\rm ext}}

\def\br{{\bf r}}

\begin{document}
\title{Kinetic and interaction components of the exact time-dependent correlation potential}
\author{Kai Luo}
\author{Johanna I. Fuks}
\author{Ernesto D. Sandoval}
\affiliation{Department of Physics and Astronomy, Hunter College and the Graduate Center of the City University of New York, 695 Park Avenue, New York, New York 10065, USA}
\author{Peter Elliott}
\affiliation{Max-Planck-Institut f\"{u}r Mikrostrukturphysik, Weinberg 2, 06120 Halle (Saale), Germany}
\author{Neepa T. Maitra}
\email{nmaitra@hunter.cuny.edu}
\affiliation{Department of Physics and Astronomy, Hunter College and the Graduate Center of the City University of New York, 695 Park Avenue, New York, New York 10065, USA}
\date{\today}

\begin{abstract}
The exact exchange-correlation (xc) potential of time-dependent density
functional theory 
has been shown to have striking features. For example, step and
peak features are generically found when the system is far from its
ground-state, and these depend nonlocally on the density in space and
time.
 We analyze the xc potential by decomposing it into 
kinetic and interaction components and
comparing each with their exact-adiabatic counterparts, for a range of
dynamical situations in model one-dimensional two-electron
systems. 
We find that often, but not
always, the kinetic contribution is largely responsible for these features, that are missed by the adiabatic approximation.  The
adiabatic approximation often makes a smaller error for the interaction
component, which we write in two parts, one being the Coulomb
potential due to the time-dependent xc hole. Non-adiabatic features of the kinetic component were also larger than those of the interaction component in
 cases that we studied when there is  negligible step structure.  In ground-state situations, step and peak structures
arise in cases of static correlation, when more than one determinant
is essential to describe the interacting state. We investigate the
time-dependent natural orbital occupation numbers and find the
corresponding relation between these and the dynamical step is more
complex than for the ground-state case.
\end{abstract}
\maketitle 

\section{Introduction}
Despite significant success in obtaining excitation spectra and
response of molecules and solids, the reliability of time-dependent
density functional theory (TDDFT)~\cite{RG84,TDDFTbook,Carstenbook}
for dynamics beyond the perturbative regime remains somewhat
cloudy. TDDFT is today increasingly stepping into the fascinating
playground of time-resolved dynamics in the presence of external
fields, and has already proven to have made useful predictions for a
number of phenomena, e.g. coherent phonon generation~\cite{SSYI12},
photovoltaic design~\cite{Rozzi13,PDP09}, dynamics of molecules in
 strong laser fields~\cite{BKPB11}, including coupling to
ions~\cite{BV12}, and attosecond control~\cite{C13}.  For many of
these applications, there is no other practical theoretical method
available that captures correlated electron dynamics for systems of
these sizes.  Although in theory exact, the reliability of TDDFT in
practise depends on the accuracy of the available approximations for
the exchange-correlation (xc) functional.  Comparison with experiment, when
it can be done meaningfully, shows that TDDFT often gets in the
ballpark but, not always, and that there is a need to understand where the errors are in the approximations, and to develop improved
approximations.

The key player in real-time TDDFT calculations is the
xc potential, $v\xc[n;\Psi_0,\Phi_0](\br,t)$, a
functional of the time-dependent one-body density $n(\br,t'<t)$, the initial interacting state
$\Psi_0$, and the initial Kohn-Sham (KS) state $\Phi_0$. Almost
all calculations today use an adiabatic approximation, which
inputs the instantaneous density into a chosen ground-state
approximation: $v\xc^{\rm adia}[n;\Psi_0,\Phi_0](\br,t) = v\xc^{\rm
  g.s.}[n(t)](\br,t)$, neglecting all memory-dependence~\cite{TDDFTbook}. Indeed calculations with such adiabatic
approximations have propelled TDDFT forward in the linear response
regime, and users generally are aware to be cautious in interpreting
their results for excitations for which the adiabatic approximation is
known not to work (e.g. multiple excitations, long-range charge
transfer between open-shell fragments, excitonic Rydberg series in
solids...)~\cite{TDDFTbook,Carstenbook}. In some cases hybrid functionals are used, which mix in a fraction of Hartree-Fock exchange, and, via their orbital-dependence, these capture some memory-dependence and non-local spatial-dependence, while still treating correlation adiabatically. 
Little is known about the performance of adiabatic functionals for non-perturbative dynamics, even for systems where the adiabatic approximation is known to perform satisfactorily within the linear response regime.  Beyond the linear response realm one must
consider the full time-dependent xc potential, not just perturbations
of it around the ground-state. To this end, there has recently been
considerable effort in finding {\it exact} xc potentials for non-equilibrium
dynamics~\cite{TGK08,RG12,EFRM12,FERM13,RNL13}, with the hope that
analysis and understanding of their main features would lead to understanding errors in the commonly used approximations, and eventually to the development of
improved functional approximations.

About 25 years ago in {\it ground-state} density-functional theory,
decompositions of the exact ground-state xc potential into kinetic and
interaction (hole) and response components began to be
considered~\cite{BBS89,GLB94,GLB96,GB96}, for the purpose of analysis
of the xc potential in cases where it could be calculated exactly, or
highly accurately.  It was found that the  component due to the Coulomb
potential of the xc hole tends to be important in real atoms and
molecules in most regions, while the kinetic and response components
play more of a role in intershell and bonding regions especially for ``stretched'' molecules, displaying step and peak features .

In the present paper we perform a similar
decomposition for the time-dependent xc potential,
particularly with a view to appraise the performance of the adiabatic
approximation. We ask,  
can a decomposition into kinetic and interaction contributions in the time-domain
provide us with insight and understanding of the
time-dependent xc potential? Recent work~\cite{EFRM12,FERM13} has shown
the prevalence of dynamical step features in the correlation potential in non-linear dynamics that require non-local
dependence on the density in both space and time; these features
appear far more generically than in the ground-state case, and are not
associated with fractional charge prevention, ionization, or electric fields,
as has been the case with steps found previously in time-dependent xc potentials. The physics of the time-dependent screening that the step
feature, and accompanying peak, represent, have yet to be understood,
and motivates the present study. Which terms in the 
decomposition of $v\xc(t)$ are largely responsible for their appearance? 
 Although it has been
shown that an adiabatic approximation completely misses the dynamical
step feature -- even in an {\it adiabatically-exact} approximation where the exact ground-state potential is used
adiabatically -- are adiabatic approximations to any of the individual
components in the time-dependent decomposition adequate?  
In the ground-state, the step
structure is a signature of static correlation, and we ask whether
this is true also for the dynamical step.  That is, is the dynamical
step an indication that the system is evolving ``significantly away''
from a single-Slater determinant (SSD)? To this end, we investigate
the dynamics of the time-dependent natural orbital occupation numbers
(NOONs) of the interacting spin-summed density-matrix. 
More generally, we will use the decomposition to try to gain a better understanding of time-dependent correlation, steps or no steps.  For example, when the system is in an excited state, there is large non-adiabatic correlation: is the kinetic or interaction component largely responsible for this?
How do the kinetic and interaction components look in cases where the density of the $N$-electron system  is a sum of $N$ spatially-separated time-evolving one-electron densities?

Section~\ref{sec:decomp} presents the decomposition of the xc
potential into the kinetic and interaction contributions; the latter we break further into two terms, one of which is the Coulomb potential due to the time-dependent xc hole. We briefly
discuss a ground-state example, and define the NOONs in Section~\ref{sec:NOs}.  In Section~\ref{sec:2eexamples} we
begin by introducing the systems and dynamics under investigation in
this paper. We focus on two-electron systems in one-dimension (1D) for
which numerically exact solutions to the dynamics are straightforward
to obtain. 
To make the problem even simpler numerically we consider
dynamical processes that involve essentially only two interacting
states: at any time  a projection onto eigenstates of the unperturbed interacting system is appreciable only for two states during the time-evolution. 
We study three cases: resonant Rabi oscillations induced by an
electric field between the ground and lowest singlet excited state in
a 1D model of the helium atom, field-free oscillations of a
superposition state in the same system, and resonant excitation energy
transfer in a 1D model of the hydrogen molecule. In each case we
plot the exact kinetic and interaction components of the correlation potential,
and compare with the adiabatically-exact
approximation.
We also
compute the time-dependent NOONs for each case, and explore their
connection with the dynamical step features. Finally, in
Section~\ref{sec:conclusions}, we briefly summarize.
 
\section{Decomposition of the XC Potential}
\label{sec:decomp}
The ground-state decomposition of the xc potential explored in
Refs.~\cite{BBS89,GLB94,GLB96,GB96} was derived from taking functional
derivatives of the kinetic and interaction contributions to the xc
energy. In the time-dependent case, we instead consider equations of motion for the current-density and density of the interacting and KS systems. 
We have~\cite{RG84,TDDFTbook,Carstenbook,MB01,L99,RB09b}
\ben
\ddot{n}(\br,t) = \nabla\cdot\left(n\nabla v\ext\right) +i\nabla\cdot\langle\Psi(t)\vert [\hat{j}(\br),\hat{T} +\hat{W}]\vert\Psi(t)\rangle
\label{eq:ndoubledot}
\een
for the interacting system, evolving under Hamiltonian $\hat{H} =
\hat{T} + \hat{W} + \sum_i^N v\ext(\hat{r_i},t)$, where $\hat{T}$ and $\hat{W}$
are the kinetic and electron-electron interaction operators
respectively.  Atomic units are used throughout this paper, $m_e = e^2
= \hbar = 1$. (We have omitted most variable-dependence on the
right-hand-side to avoid notational clutter).  A similar equation
holds for the KS system where the KS Hamiltonian has $\hat{W}=0$ and
the external potential $v\ext$ is replaced by the KS potential $v\s =
v\ext + v\H + v\xc$, the sum of the external, Hartree, and xc terms.
Since the KS system evolves with identical density to the interacting
system, we equate the right-hand-sides of Eq.~(\ref{eq:ndoubledot}) and
its KS analog, to find
\begin{widetext}
\ben
\nabla\cdot\left(n\nabla v\xc\right) =\nabla \cdot \left[ \frac{1}{4}\left(\nabla' - \nabla\right)\left(\nabla^2 - \nabla'^2\right) \left( \rho_1(\br',\br,t) - \rho_{1,s}(\br',\br,t) \right) \vert_{\br'=\br} + n(\br,t)\int n\xc(\br',\br,t) \nabla w(\vert \br'-\br \vert) d^3r'\right] ,
\label{eq:3Dvxc}
\een
where 
$
\rho_1(\br',\br,t) = N\sum_{\sigma_1..\sigma_N} \!\!\int d^3r_2...d^3r_N \Psi^*(\br'\sigma_1,\br_2\sigma_2...\br_N\sigma_N;t) \Psi(\br\sigma_1,\br_2\sigma_2 \dots \br_N\sigma_N;t)$
is the spin-summed one-body density-matrix of the true system of electrons with two-body interaction potential $w(\vert\br - \br'\vert)$, $\rho_{1,\sss{S}}(\br',\br,t)$ is the one-body density-matrix for the Kohn-Sham system,  and $n\xc(\br',\br,t)$ is the xc hole, defined via the pair density, 
$P(\br',\br,t) = N(N-1)\sum_{\sigma_1..\sigma_N}\int \vert \Psi(\br'\sigma_1,\br\sigma_2,\br_3\sigma_3..\br_N\sigma_N; t) \vert^2 d^3r_3..d^3r_N 
 = n(\br,t)\left(n(\br',t) +n\xc(\br',\br,t)\right)$\;.
\end{widetext}
Eq.~(\ref{eq:3Dvxc}) is a
Sturm-Liouville equation for $v\xc$, giving a unique solution for a
given density $n(\br,t)$ and boundary condition~\cite{L99}.
The first term in Eq.~(\ref{eq:3Dvxc}) gives a kinetic-like contribution to the xc potential while the second term is a contribution stemming directly from the electron-electron interaction that depends on the xc hole. 
In 1D, Eq.~(\ref{eq:3Dvxc}) can be easily solved for the xc field, defined as the gradient of the xc potential:
\begin{widetext}
\begin{equation}
\frac{d}{dx} v\xc(x,t) = \frac{1}{4n(x,t)}\left(\frac{d}{dx'} - \frac{d}{dx} \right)\left(\frac{d^2}{dx^2}- \frac{d^2}{dx'^2}\right)\left(\rho_1(x',x,t) - \rho_{1,S}(x',x,t) \right)\vert_{x'=x} + \int n\xc(x',x,t) \frac{\partial}{\partial x} w(\vert x'-x \vert) dx'.
\label{eq:vxc}
\end{equation}
Note that in going from Eq.~(\ref{eq:3Dvxc}) to Eq.~(\ref{eq:vxc}), we have thrown away a term
of the form $g(t)/n(x,t)$, where $g(t)$ is the integration constant of
the outer $\nabla$ in Eq.(2). We do so because $g(t)$
is actually zero, due to satisfaction of boundary conditions: at the
boundary of a finite system, the density decays exponentially, so to
avoid the field $\nabla v\xc$ diverging exponentially, the integration
constant $g(t)$ must be taken to be zero. We observe that, unlike in
3D where the KS and true currents may differ by a rotational
component, in 1D the KS current equals the true current for finite
systems, as follows from the equation of continuity.
We now write $v\xc(x,t) = v\c^{T}(x,t) + v\xc^{W}(x,t)$ and define the kinetic contribution $v\c^T$ from the first term on the right of Eq.~(\ref{eq:vxc}):
\ben
v\c^T(x,t) \equiv \int^x \frac{1}{4n(x'',t)}\left(\frac{d}{dx'} - \frac{d}{dx''} \right)\left(\frac{d^2}{dx''^2}- \frac{d^2}{dx'^2}\right)\left(\rho_1(x',x'',t) - \rho_{1,S}(x',x'',t) \right)\vert_{x'=x''} dx''\;,
\label{eq:vcT}
\een
\end{widetext}
since it arises from differences in kinetic/momentum aspects
of the KS and interacting systems. Further, we denote it as a
correlation contribution (hence the c subscript), since correlation generally refers to the deviation from single-Slater determinant behavior.
The second term in Eq.~(\ref{eq:vxc}) gives a contribution arising directly from the electron-interaction $W$, which we denote  $v\xc^W(\br,t)$. We further decompose $v\xc^W$ as:
\ben
v\xc^W(x,t) = v\xc^{\rm hole}(x,t) + \Delta v\xc^{W}(x,t)
\label{eq:vxcWdecomp}
\een
where $v\xc^{\rm hole}$ is the Coulomb potential of the xc hole,
\ben
v\xc^{\rm hole}(x,t) = \int_{-\infty}^{\infty} dx' \; n\xc(x',x,t) w(\vert x -x'\vert)
\label{eq:vxchole}
\een
while the remaining term, $\Delta v\xc^{W}$, is  
\ben
\Delta v\xc^W(x,t)= -\int^x dx''\int_{-\infty}^{\infty} dx' w(\vert x'' -x'\vert)\frac{\partial}{\partial x''} n\xc(x',x'',t)\;,
\label{eq:DeltavxcW}
\een 
where we take the lower limit of
the $x''$-integrations in Eq.~(\ref{eq:vcT}) and Eq.~(\ref{eq:DeltavxcW})  as zero in our calculations. A different
choice simply shifts the potential uniformly by an irrelevant spatial
constant.

Before proceeding, we consider a simple example.  Consider a system of
{\it non-interacting} electrons evolving from an initial state
$\Psi_0$ in some potential $v(x,t)$. We may then ask whether we can
find a potential in which the same non-interacting electrons evolve
with exactly the same density but beginning in a different initial
state $\Phi_0$~\cite{MB01,EM12}. Assuming such a potential may be
found, we see that the potential that the second system evolves in is
given by $v(x,t) + v\c^T(x,t)$. That is the kinetic part of the
potential contains the entire difference. From this simple argument,
we might expect that $v\c^T$ in the general interacting case contains
a large part of the initial-state dependent effects. In fact, in our
examples that do not start from the ground-state, we shall see $v\c^T$
is indeed the predominant term in the initial correlation potential.

Returning to the decomposition, a similar decomposition in the
ground-state has led to insights for ground-state potentials in
various cases~\cite{BBS89,GLB94,GLB96,GB96,TMM09}. There, the exact ground-state xc
potential is decomposed into
a 
 kinetic contribution $v\c^{\rm kin}$, the Coulomb potential due
to the xc hole $v\xc^{\rm hole}$, and two response terms that depend on the functional derivatives of these two potentials with respect to the density, denoted together as $v\xc^{\rm resp}$; namely, 
$v\xc(\br)= v\c^{\rm kin}(\br) + v\xc^{\rm hole}(\br) +v\xc^{\rm  resp}(\br)$. 
In real atoms and molecules at
equilibrium, it is expected that $v\xc^{\rm hole}$ is the important
contribution to $v\xc$ in most regions, as demonstrated in
Refs.~\cite{BBS89,GLB94,GLB96,GB96}. The kinetic potential tends to
give peaks in intershell regions in atoms and bonding regions in
molecules, while the response potential may have step structures
related to different decays of the dominant orbitals. These steps and
peaks do however become more prominent in molecules stretched to large
bond-lengths and are associated with static correlation. 
(A note if we wish to compare this decomposition with the time-dependent one presented here when applied to ground-states: although the hole potential $v\xc^{\rm hole}$ of Eq.~(\ref{eq:vxchole}) reduces to the $v\xc^{\rm hole}$ of the ground-state decomposition, 
$v\c^{T}$ does not quite reduce to $v\c^{\rm kin}$, since $v\c^{T}$ also includes part of $v\c^{\rm resp}$. Likewise, $\Delta v\c^{W}$ would then reduce to the remaining part of $v\c^{\rm resp}$.)

\begin{figure}[h]
\begin{center}
\includegraphics[width=0.45\textwidth]{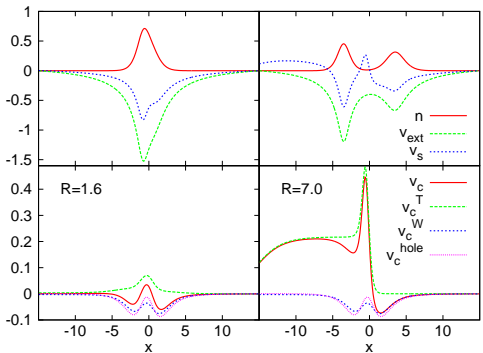}
\caption{Ground-state potential components for a 1D model of the LiH molecule~\cite{TMM09} for equilibrium $R=1.6$~a.u. (left) and stretched $R=7.0$~a.u. (right) geometries. Top panels: density (red solid), external potential (green dashed) and Kohn-Sham potential (blue dotted). Lower panels: $v\c^W$ (blue dotted), $v\c^T$ (green dashed) and $v\c^{\rm hole}$ (pink dotted) contributions to the total correlation potential $v\c$ (red solid). 
}
\end{center}
\label{fig:1dLiHgs}
\end{figure}

An example of
a 1D model of a LiH molecule is shown in Figure~\ref{fig:1dLiHgs}, where
two fermions, interacting via $1/\sqrt{1+(x_1-x_2)^2}$ live in the
potential $v\ext(x) = -1/\sqrt{2.25 +(x+R/2)^2} - 1/\sqrt{0.7
  +(x-R/2)^2}$ (see Ref~\cite{TMM09} for details). Moving from
equilibrium separation of $R=1.6$au to larger bond lengths, a salient
feature is the build-up of the step and peak structures in
$v\c^T$. These features are essential to prevent dissociation of the
molecule into fractional charges, and to lead to the correct
atomic-densities in the infinite separation limit.  The kinetic
component $v\c^T$ gives the correlation potential
an ultra-non-local in space character, while the hole potential,
$v\c^{\rm hole}$ is quite local~\cite{GB96,TMM09}. 
In the general case, approximations in use today do a
better job of capturing $v\c^{\rm hole}$ than of $v\c^{T}$ and
$\Delta v\c^{W}$, which require the correlation potential to have
spatially non-local density-dependence.  

In the present paper, we explore the decomposition represented in
Eqs.~(\ref{eq:vxc}) -- (\ref{eq:DeltavxcW}), with the hope of gaining
insight and understanding of the time-dependent xc potentials, as
described in the introduction. We focus on the correlation potential here since we will consider two-electron spin-singlet systems, taking the KS state as a doubly-occupied orbital: in this case, the exchange-potential is simply minus half the Hartree-potential, $v\x(x,t) = - v\H(x,t)/2$, and the exchange-hole is minus half the density, $n\x(x,t)= -n(x,t)/2$. 
 One focus will be on the dynamical step
and peak structures found in the earlier works of
Refs.~\cite{EFRM12,FERM13}. Quite generally, time-dependent step and
peak features were found in the time-dependent correlation potential
of two-electron systems, for dynamics beyond the linear response
regime~\cite{LEM13}, that cannot be captured by any adiabatic
approximation. Having non-local density-dependence in space and in
time, they are a challenge to incorporate in functional
approximations, but their absence might have a significant effect on the
dynamics. We examine the kinetic and hole contributions to the
correlation potential to try to gain a better understanding of the
time-dependent screening these features represent; whether the
screening is largely due to kinetic or interaction effects.  We already
notice that such features do not appear in the hole component
$v\xc^{\rm hole}$: taking $x$ large in Eq.~\ref{eq:vxchole} shows that
asymptotically  far from the system $v\xc^{\rm hole} \to -1/x$, discarding
the possibility of a dynamical step across the system in this component. 

We will investigate whether the adiabatically-exact approximation (see
Sec.~\ref{sec:2eexamples}) is adequate for any of the components
$v\c^{T}, v\c^W, v\c^{\rm hole}$: this will indicate the ``best'' an
adiabatic functional can do. It is not just the step structures we are interested in: we will also consider what the different components of the correlation potential and their adiabatic counterparts look like when no noticeable dynamical step is present, e.g. when the system is in an excited state (one of the time snapshots in Sec.~\ref{sec:sftcHe-Rabi}), and in a case where throughout the dynamics no noticeable step features are observed (Sec.~\ref{sec:RET}).
In the latter case, the
system consists of widely-separated atoms, each with time-evolving
one-electron densities.  Perhaps surprisingly, the exact correlation potential  shows large features in the one-electron regions, that are completely
missed by the adiabatic approximation. These features appear not only in
regions of negligible density between the atoms, as has
been found in the ground-state (e.g. Fig.~\ref{fig:1dLiHgs} above),
but actually in the regions where each electron lives.  We show that
$v\c^T$ is responsible for these features and discuss why.

\subsection{Natural orbitals and steps in the correlation potential}
\label{sec:NOs}
Another aspect of the dynamics we will investigate is the relation
between the dynamical step structures and the time-dependent
NOONs. The NOONs (defined shortly) are eigenvalues of the spin-summed
one-body reduced density matrix, and take on values between 0 and
2. For a SSD, each NOON is either 2 or 0. 
 The step structure in the
{\it ground-state} potential indicates strong correlation in the system,
with NOONs significantly away from their SSD values.  For example, the
largest occupation numbers in the equilibrium geometry in the model of
the LiH molecule in Fig.~\ref{fig:1dLiHgs}  are 1.9551, 0.0412, 0.0035...,
indicating a weakly correlated system, while for the stretched molecule at $R=7$au they are
1.0996, 0.8996, 0.0008... As the separation increases further, the two
largest occupation numbers approach one, with all others becoming zero. This
indicates a strong deviation from SSD behavior. 

By
diagonalizing the one-body time-dependent density-matrix of the
interacting system, $\rho_1(x,x',t)$, we will investigate the connection between the time-dependent
NOONs and the dynamical step. 
In each example, we  will diagonalize
the interacting $\rho_1$:
\ben
\int \rho_1(x,x',t) \varphi^*_j(x',t) d x' = \eta_j(t) \varphi_j(x,t) 
\een
The eigenfunctions $\varphi_j$ are called natural orbitals (NOs) and the eigenvalues $\eta_j$ are the NOONs.
In Ref.~\cite{EFRM12} it was argued that, in the two-electron case,
the step structures appear at peaks of the acceleration, with
magnitude given by the spatial integral of the acceleration: in the
expression for the KS potential, there is a term
$\int^x\partial_t(j(x',t)/n(x',t)) dx'$, where $j(x,t)$ is the one-body current-density,  which is responsible for the
dynamical step.  It is straightforward to show that in the general
$N$-electron case, 
\ben
\begin{split}
\partial_t \left({\frac{j(x,t)}{n(x,t)}}\right) = & \sum_k{\eta_k}(t)\left(\frac{\partial_t j_k(x,t)}{n(x,t)} - \frac{j(x,t)}{n^2(x,t)} \partial_t n_k(x,t)\right)
\\ & +\sum_k \dot{\eta}_k(t)\left(\frac{j_k(x,t)}{n(x,t)} -\frac{j(x,t)}{n^2(x,t)} n_k(x,t)\right)
\label{eq:step-NO}
\end{split}
\een
where
\ben
n_k(x,t) = |\varphi_k(x,t)|^2,\; {\rm and}
\een
\ben
j_k(x,t)=\frac{-i}{2}\left[ \varphi^*_k(x,t)\nabla \varphi_k(x,t) - \varphi_k(x,t) \nabla \varphi^*_k(x,t) \right].
\een
The spatial integral of the right-hand-side of Eq.~(\ref{eq:step-NO})
gives the dynamical step structure studied in Ref.~\cite{EFRM12} expressed in
terms of time-dependent NOs and NOONs.
The relation is far from trivial, and suggests that the relation
between the dynamical step and the time-dependent NOONs is not as
straightforward as that between the ground-state step structures and
the ground-state NOONs. We will plot the NOONs $\eta_k(t)$ for the different dynamics presented in this work, and see if any trends can be identified.

\section{Results: Dynamics of Two Electrons in One-Dimension}
\label{sec:2eexamples}
In order to find the exact xc potential Eq.~(\ref{eq:vxc}), we must not
only solve for an exact, or highly accurate, solution for the
interacting wavefunction, from which we extract $\rho_1(x,x',t)$ and $n\xc(x,x',t)$,
but we also need a method to find the exact KS density-matrix
$\rho_{1,\sss{S}}(x,x',t)$.  In general this calls for an iterative
scheme~\cite{RG12,RNL12}, but for two electrons in a singlet state,
assuming one starts the Kohn-Sham calculation in a single
Slater-determinant, then simply requiring the doubly-occupied KS orbital to reproduce the exact density $n(x,t)$ of the interaction problem, yields
\begin{equation}
\phi(x,t) = \sqrt{\frac{n(x,t)}{2}} \: e^{i\int^x \frac{j(x',t)}{n(x't)}dx'}
	\label{eq:rhoks}
\end{equation}
and 	$\rho_{1,\sss{S}} (x',x,t) = 2 \: \phi^*(x',t)\phi(x,t)$.

We note the alternate way of finding the correlation potential in
Ref.~\cite{EFRM12}; there $v\s (x,t)$ is found first, by choosing the
initial KS state to be a doubly-occupied spatial orbital and inverting
the KS equations (Eq (1) of Ref.~\cite{EFRM12}).  Then $v\xc(x,t)$ is
obtained by subtracting the Hartree potential $v\H(x,t)$ and the
external potential $v\ext(x,t)$ at time $t$ (Eq. 2 of
Ref.~\cite{EFRM12}).  In the present approach, we instead extract the
xc potential directly from Eq.~(\ref{eq:vxc}). There are two advantages:
the first, is that Eq.~(\ref{eq:vxc}) is valid for $N$-electrons (and
it's precursor Eq.~(\ref{eq:3Dvxc}) is valid also for three dimensions),
while the expression for the KS potential used in Ref.~\cite{EFRM12}
is only valid for two electrons. The second, is that since it is an
expression for $v\xc$ explicitly, it more readily points to what
functional approximations must approximate: the right-hand-side of Eq.~(\ref{eq:vxc}) is
what needs to be approximated as a functional of the density (see also Ref.~\cite{RB09b}). On the
other hand, the expression in Ref.~\cite{EFRM12} for the xc potential
has terms between the KS potential and the external potential that
cancel in a subtle hidden way, and it is harder to see what terms the
xc potential should be aiming to approximate.

In the following we consider various two-electron dynamics that either
start in the ground-state and evolve far from it, or begin in a
non-stationary state. Our 1D ``electrons'' interact via the soft-Coulomb interaction $w(x',x) =1/\sqrt{(x'-x)^2+1}$ and 
live in either a 1D atom (sections \ref{sec:sftcHe-Rabi}-\ref{sec:field-free}) or a 1D molecule (Section~\ref{sec:RET}).

In our examples, the interacting dynamics largely, if not fully,
involve two interacting states. This means that we can solve
for the time-dependent interacting wavefunction $\Psi(t)$ in a particularly
straightforward manner. Assuming a two-state Hilbert space,
\ben
\label{2lvlPSI}
 |\Psi(t)\rangle = a_1(t) |\Psi_1\rangle  + a_2(t) |\Psi_2\rangle\;,
\een 
then for the field-free cases (sections \ref{sec:field-free},\ref{sec:RET}) the time-dependent coefficients are simply given by $a_j(t) = e^{-i E_j t}$,
where $E_j$ is the eigenvalue of state $\Psi_j$.  For dynamics in a resonant external field (section \ref{sec:sftcHe-Rabi}) the only two states involved in the dynamics are the ground state $\Psi_g$ and  the first dipole-allowed excited state $\Psi_e$,  and $a_g(t), a_e(t)$ are solutions of the two-level Schr\"odinger equation,
\ben
i \partial_t  \left( {\begin{array}{c}
 a_g(t)\\
 a_e(t)  \\
 \end{array} } \right)=
\left( {\begin{array}{cc}
 E_g -d_{gg}{\cal E}(t)  &  -d_{eg}{\cal E}(t)  \\
-d_{eg}{\cal E}(t) & E_e -d_{ee}{\cal E}(t)  \\
 \end{array} } \right) \left( {\begin{array}{c}
 a_g(t)   \\
 a_e(t)
 \end{array} } \right)
\label{coeffsEOM}
\een
where $E_g$, $E_e$ are the energy eigenvalues of the two
states, $d_{ab} = \int \Psi_a^*(x_1,x_2)(x_1+x_2)\Psi_b(x_1,x_2)
dx_1dx_2 $ is the transition dipole moment and ${\cal E}(t) =
A\cos(\omega t) $ is an applied electric field of strength $A$ and
frequency $\omega$.   For $\omega \gg |d_{eg}A| $ and $\omega$ close to the resonant frequency, this reduces to the textbook Rabi problem; 
the period of the oscillations between the ground and excited state for a resonant applied field is given by $T_R = \frac{2 \pi}{ \vert d_{eg} A \vert }$, in the case where the ground and excited state each have a zero dipole moment, $d_{gg} = d_{ee} = 0$. 

We have compared the results from the two-state approximation with a
full real-space calculation, solving the exact time-dependent Schr\"odinger equation using the octopus code~\cite{octopus, octopus2}; 
aside from asymptotic features, they largely agree. We note that in the low-density region far from
the system, the potential is unreliable due to noise, however this
does not affect the region of interest shown in the figures. As we move out
further from the atomic/molecular center, higher-excited states that
are neglected in the two-state approximation come into play. These
states contribute to polarization of the density, especially
asymptotically where the contribution of the two lower energy states has dropped due to their faster decay.
In the two-state approximation, this polarization effect is missing.  When
the correlation potential is extracted from the total KS potential
$v\s$ as was done in Ref.~\cite{EFRM12} a field-counteracting term
appears in the correlation potential $v\c$ to counter the external
field in $v\ext$: because of the absence of polarization within the two-state model, the KS
potential generated using information of the density and
current-density in Ref.~\cite{EFRM12} must be flat (i.e. constant)
asymptotically. However in the present approach, $v\c$ is generated
directly from Eq.~(\ref{eq:vxc}) where the input density-matrix and xc
hole are computed within the two-state approximation and so lack
asymptotic polarization. This means that no field-counteracting term
in the correlation potential is present in the present approach,
similar to the full real-space case.


In all calculations we compare with the {\it adiabatically-exact} (AE)
approximation: $v\c^{\rm AE}[n;\Psi_0,\Phi_0](\br,t) = v\c^{\rm exact-
  g.s.}[n(t)](\br,t)$. Note that the AE approximation for exchange
coincides with the exact exchange potential for two electrons, since
in this case $v\x = -v\H/2$ has only instantaneous dependence on the
density.  The AE approximation takes both the underlying interacting
and KS wavefunctions at time $t$ to be ground-state wavefunctions of
density equal to the true density at time $t$.  To find the AE
correlation potential, we first find the ground-state interacting
wavefunction of density $n(t)$, $\Psi^{\rm g.s.}[n(t)]$, using the
iterative scheme of Refs.~\cite{TGK08,EFRM12}, and the ground-state KS
wavefunction which is the doubly-occupied orbital
$\sqrt{n(t)/2}$. From these, we find the reduced quantities
$\rho_1^{\rm AE}, \rho_{1,\sss{S}}^{\rm AE}$ and $n\xc^{\rm AE}$ to insert into Eq.~(\ref{eq:vxc}).

\subsection{1D He: Rabi dynamics to local excitations}
\label{sec:sftcHe-Rabi}
\begin{figure}[h]
\begin{center}
\includegraphics[width=0.45 \textwidth]{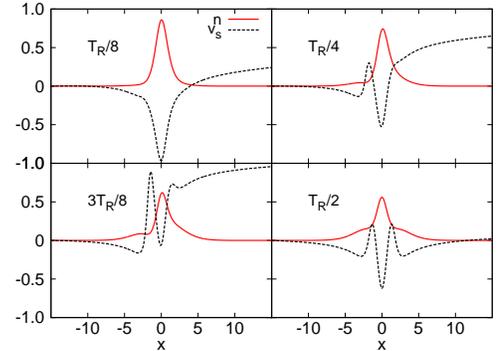}
\end{center}
\caption{(Color online) 1D He model: 
snapshots of density $n$ (red solid) and exact KS potential $v\s$ (black dashed) during a half-Rabi cycle (excited state is reached at $T_R/2$).}
\label{fig:sftcHe-n_vs}
\end{figure}

Here we consider a 1D model of the He atom $v\ext(x,0) =
-2/\sqrt{x^2 +1}$, and apply a weak resonant field ${\cal E}(t) = 0.00667 \cos(0.533 t)$ to induce local Rabi oscillations between the ground and the lowest singlet excited state of the system. (The Rabi frequency is  $d_{eg}A=0.00735$~au).
This system was also considered in Refs.~\cite{RB09,FHTR11,EFRM12}. 
Figure~\ref{fig:sftcHe-n_vs} plots the exact KS potential at four
times during a half-Rabi cycle, along with the density. Step and peak
structures are clearly present during the time-evolution. The step
actually oscillates on the time-scale of the optical cycle, with magnitude and direction varying significantly, as
evident in Figure~\ref{fig:sftcHe_tdnos_Topt}, where snapshots over an optical time
slice near $T_R/4$ are shown. (We comment on  this figure later).   

\begin{figure}[h]
\begin{center}
\includegraphics[width=0.45 \textwidth]{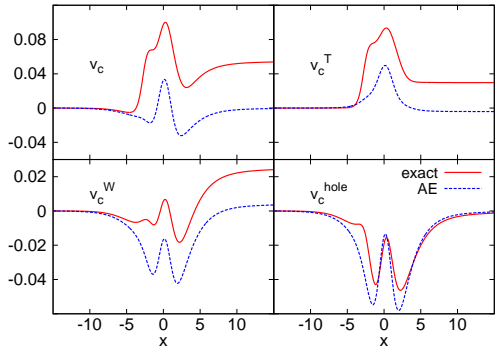}
\end{center}
\caption{(Color online) 1D He Rabi dynamics at $T_R/8$: exact (red solid) and AE (blue dashed) components of $v\c$ as indicated. 
}
\label{fig:sftc_T1o8}
\end{figure}

\begin{figure}[h]
\begin{center}
\includegraphics[width=0.45 \textwidth]{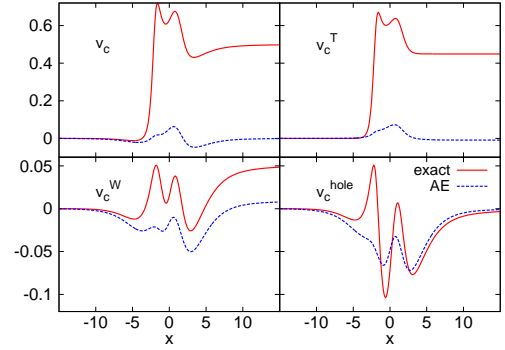}
\end{center}
\caption{(Color online) 1D He Rabi dynamics at $T_R/4$ (see caption Figure~\ref{fig:sftc_T1o8})} 
\label{fig:sftc_T1o4}
\end{figure}

\begin{figure}[h]
\begin{center}
\includegraphics[width=0.45 \textwidth]{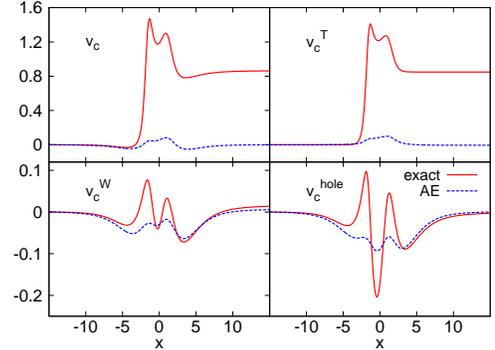}
\end{center}
\caption{(Color online) 1D He Rabi dynamics at $3T_R/8$.} 
\label{fig:sftc_T3o8}
\end{figure}

The correlation potential $v\c$ is responsible for these dynamical steps,
as discussed in Ref.~\cite{EFRM12}, and now we investigate the role of
the different components $v\c^T,v\c^W$, and $v\c^{\rm
  hole}$. Figure~\ref{fig:sftc_T1o8} compares these components with
their AE approximations, at $T_R/8$. As was noted in
Ref.~\cite{EFRM12}, the AE approximation does not capture the
dynamical step at all, however what we find here (top right and lower
left panels) is that both $v\c^{T,{\rm AE}}$ and $v\c^{W,{\rm
    AE}}$ do display a small step feature, that exactly cancel once
added. Although $v\c^{{\rm AE}}$ does a poor job in approximating
$v\c$, the AE approximation is noticeably better for the hole
component: $v\c^{\rm hole, AE}$ does somewhat capture $v\c^{\rm hole}$ as shown in the lower right panel, reasonably capturing the well structure. Neither the exact nor the AE $v\c^{\rm hole}$ component  displays any step structure. 
These observations appeared to hold generally; for example, see Figure~\ref{fig:sftc_T1o4}, where the components are shown at $T_R/4$. There, the step is considerably larger than at $T_R/8$, and the dominant component to the step appears in $v\c^T$, while at $T_R/8$, the contributions from $v\c^W$ and $v\c^T$ are comparable. Again, the $v\c^T$ and $v\c^W$ components of the AE approximation each display a (much smaller) step, but which cancel each other; again the AE approximation does a better job for $v\c^{\rm hole}$ than for the other components.
Figure~\ref{fig:sftc_T3o8} shows the components at $3T_R/8$, where practically all of the step is in the kinetic component $v\c^T$; still, the AE approximation approximates none of the components well.

\begin{figure}[h]
\begin{center}
\includegraphics[width=0.45 \textwidth]{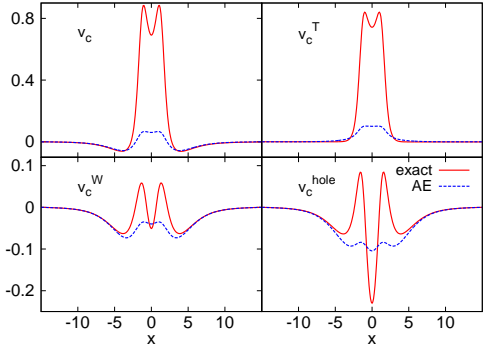}
\end{center}
\caption{
(Color online) 1D He Rabi dynamics at $T_R/2$, where the true state has reached the excited state of the system.} 
\label{fig:sftc_T1o2}
\end{figure}

At the time when the excited state is reached, the dynamical step
wanes: as $T_R/2$ is reached, the electron dynamics slows down, and
the local acceleration in the system decreases to zero (see Fig.\ref{fig:sftc_T1o2}). As was argued
in Ref.~\cite{EFRM12} the dynamical step arises from a spatial
integral of the acceleration through the system, so consequently this
goes to zero; the oscillations over the optical cycle become
increasingly gentle and eventually vanish to
zero. Figure~\ref{fig:sftc_T1o2} shows that still, the AE correlation
potential is dramatically different from the exact potential, and that
the dominant non-adiabatic features are contained in the kinetic component
$v\c^T$. The AE approximation does not do well for any of the
components, but is particularly bad for $v\c^T$. This can be
understood from realizing that underlying the AE approximation is the
assumption that both the interacting and KS states are
ground-states. This is obviously not the case at half a Rabi cycle,
when the true state has reached the first excited state of the
system. The KS state on the other hand does have a ground-state nature
(although is not the ground-state of the 1D-He potential), as it
consists of a doubly-occupied node-less wavefunction.  One can interpret
this result in terms of initial-state dependence~\cite{EM12}: if we consider the states at $T_R/2$ to be initial states for subsequent dynamics, then the exact correlation potential is $v\c(T_R/2) = v\c[n,\Psi^*,\Phi^{\rm g.s}]$ 
while an adiabatic approximation inherently assumes that the interacting wavefunction is a ground state instead of an excited state $v\c^{\rm AE}(T_R/2) = v\c[n,\Psi^{\rm g.s.},\Phi^{\rm g.s}]$.

\begin{figure}[h]
\begin{center}
\includegraphics[width=0.45 \textwidth,angle = 0]{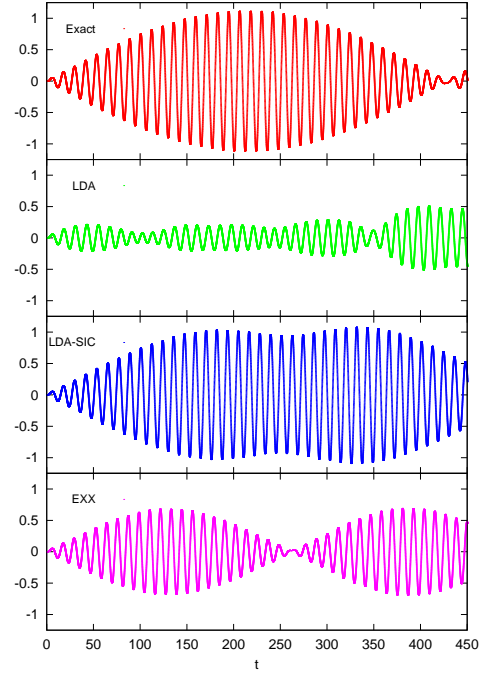}
\end{center}
\caption{
(Color online) Dipole moment $d(t) = \int n(x,t)x dx$ during a half-Rabi cycle for the 1D He model. The same field is applied in all cases, ${\cal E}(t) = 0.00667 \cos(0.533 t)$.  Exact (top panel), LDA (second panel), LDA-SIC (third panel), and EXX (fourth panel).} 
\label{fig:dipoles}
\end{figure}

Although the dynamical step structures look rather stark, they do tend
to appear in regions where the density is small, although not
negligible. A question is then, what is their impact on the dynamics?
Figure~\ref{fig:dipoles} plots the exact dipole, compared with three
TDDFT calculations using approximate functionals; in all calculations the same field is applied, resonant with the exact transition frequency.
 These approximations do quite
poorly, as has also been observed in the past for Rabi
dynamics~\cite{RB09,FHTR11}.  The linear response (LR) resonances for exact exchange(EXX), the local density
approximation (LDA) and the self-interaction corrected LDA (LDA-SIC),
 lie at $\omega^{LR}_{\rm
  EXX}=0.549$ a.u., $\omega^{LR}_{\rm LDA-SIC}=0.528$ a.u. and
$\omega^{LR}_{\rm LDA}=0.475$ a.u., whereas the exact resonance is at
$\omega=0.533$ a.u. 
Still, we note that recent work studying charge-transfer dynamics in the Hubbard model ~\cite{FM14a} and on 3D molecules ~\cite{RN11} show that even when the LR frequency of the approximation is extremely close to the exact, the non-linear adiabatic dynamics can still be poor.
The failure of the approximate methods is evident in Fig.~\ref{fig:dipoles} and is worse for the approximations
with poor LR resonances.  For LDA, in addition to the bad LR
frequency the ionization threshold  lies
already below $\omega=0.5$ a.u., so the LDA dipole begins to probe the continuum and there is no dominant frequency.  In
order to assess the impact of the adiabatic approximation itself independently
of the choice of the ground state approximation it would be
desirable to run an AE calculation
self-consistently. To this aim the iterative procedure of Ref.~\cite{TGK08}
should be performed at each time step of the propagation, which we leave to future work. 

\begin{figure}[h]
\begin{center}
\includegraphics[width=0.35\textwidth]{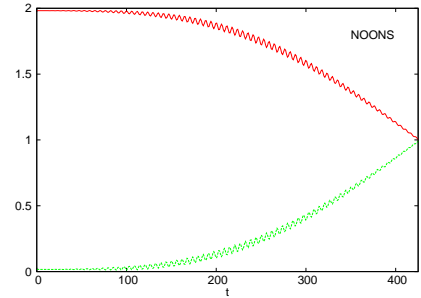}
\end{center}
\caption{(color online) The two largest time-dependent NOONs over a half-Rabi cycle for the 1D He model. All other NOONs are negligible.}
\label{fig:sftcHe_tdnos}
\end{figure}

\begin{figure}[h]
\begin{center}
\includegraphics[width=0.45 \textwidth]{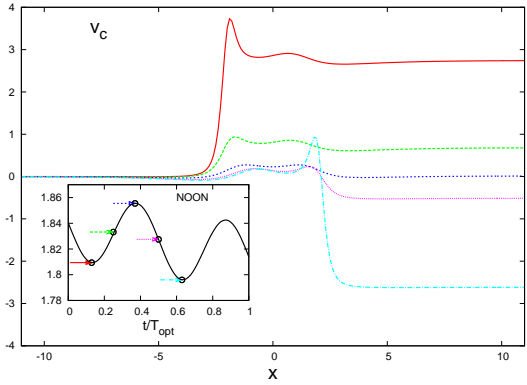}
\end{center}
\caption{(color online) 1D He Rabi dynamics: The dynamical step at snapshots over one optical cycle near $T_R/4$ , i.e. at times $0.13, 0.25, 0.38,0.5,0.63$ $T_{\rm opt}$ after $T_R/4$, as indicated in the inset; the coloured arrows indicate the corresponding times.  The dominant time-dependent NOON is shown in the inset.}
\label{fig:sftcHe_tdnos_Topt}
\end{figure}

We now come to the question of the relation between the dynamical step
and the NOONs. Figure~\ref{fig:sftcHe_tdnos} shows the NOONs plotted
over a half-Rabi cycle: as might be anticipated, two dominate. One
starts out close to 2 while the other is close to 0, and both approach
1 as the excited state is reached at $T_R/2 \approx 430$au. In particular, we note that, in
contrast with the ground-state case, there is no direct relation with
the deviation from SSD and the size of the step, e.g. as we approach a
half-Rabi cycle, when the interacting system is farthest from a SSD,
the size of the dynamical step decreases and eventually vanishes.
Instead, it seems to be related more to the local oscillatory
behavior of the NOONs: Figure~\ref{fig:sftcHe_tdnos_Topt} shows the
step at various times in an optical cycle near $T_R/4$
while the inset shows the corresponding NOONs. We observe that there
is a correlation between the oscillations of the step and those of the
NOONs. The largest(smallest) magnitude for the step size during the optical cycle appears to
occur at local minima(maxima) of NOONs. This feature also holds when we zoom
in to optical cycles centered around other times. Considering the
complexity of Eq.~(\ref{eq:step-NO}), this result is not anticipated,
and we will now turn to another example to see if the trend holds.
The adiabatic NOONs (not shown), computed from diagonalizing the one-body density matrix of the interacting ground-state wavefunction of instantaneous density $n(x,t)$, have a much smaller variation. They
begin at the exact values (1.9819, 0.0166, 0.0014...), make a gentle
dip to (1.8437, 0.0899, 0.0668...) at $T_R/2$ before rising back up: in
the AE approximation the underlying ground-state
remains weakly correlated throughout, as it is the ground-state of a
relatively localized density. 


\subsection{1D He: Field-free evolution of a non-stationary state}
\label{sec:field-free}
\begin{figure}[h]
\begin{center}
\includegraphics[width=0.45 \textwidth]{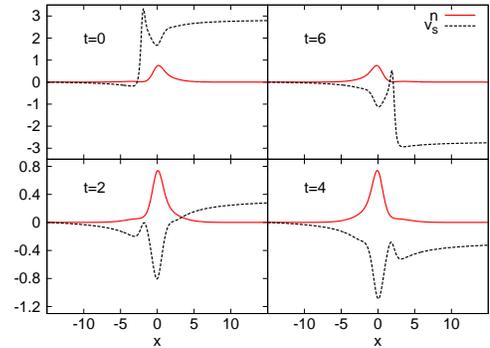}
\end{center}
\caption{The exact Kohn-Sham potential (black dashed) and density (red solid) in the field-free evolution of Eq.~(\ref{eq:IS}) in the 1D He at times indicated.}
\label{fig:FF-n_vs}
\end{figure}

\begin{figure}[h]
\begin{center}
\includegraphics[width=0.45 \textwidth]{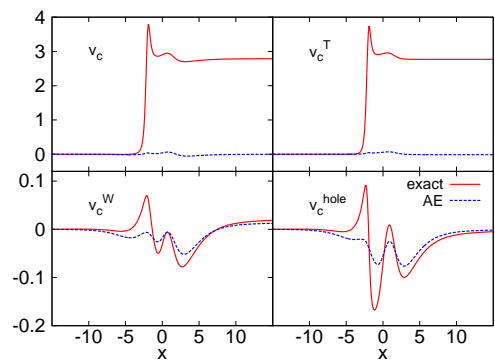}
\end{center}
\caption{(color online) Field-free evolution of Eq.~(\ref{eq:IS}) in 1D He: components of $v\c$ at the initial time.}
\label{fig:FF-vc_cptsT0}
\end{figure}

\begin{figure}[h]
\begin{center}
\includegraphics[width=0.45 \textwidth]{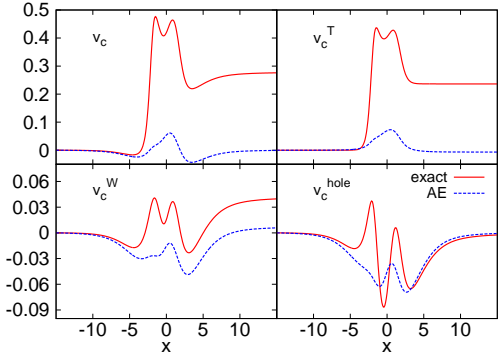}
\end{center}
\caption{(color online) As in Figure~\ref{fig:FF-vc_cptsT0} but at $t=2$au.}
\label{fig:FF-vc_cptsT2}
\end{figure}

In this example, we revisit the field-free evolution of a 50:50
mixture of the ground and first excited state presented in
Ref.~\cite{EFRM12} in the 1D He atom,
\ben
|\Psi(t)\rangle = \left(e^{-iE_g t}|\Psi_g\rangle + e^{-iE_e t}|\Psi_e\rangle \right)/\sqrt{2}\;.
\label{eq:IS}
\een
First in Figure~\ref{fig:FF-n_vs}, we plot the exact KS potential and
the density at four times  within the first half-period of the
motion (the period of the dynamics is $2\pi/(E_e - E_g) = 11.788$ au). Dynamical steps are once again clearly evident, and particularly
prominent at the initial time and every half-period of the evolution.
There it dominates the xc potential. 
Figure~\ref{fig:FF-vc_cptsT0} shows the correlation potential at the
initial time, as well as its components $v\c^T$, $v\c^W$, and
$v\c^{\rm hole}$, and the AE approximation to these
terms. We notice that the step is the over-riding feature of the
correlation potential at this time, and is largely contained in the
kinetic component $v\c^T$. This is consistent with the expectation expressed in Section~\ref{sec:decomp}, that initial-state effects are largely contained in the kinetic component of the correlation potential. 
 The AE approximation fails miserably to capture it, but does a much
better job in capturing the gentle undulations of $v\c^W$ and even more so $v\c^{\rm hole}$. The $v\c^W$ does appear to display a small step, and is
fairly captured by the AE approximation at this time.  At time $t=2$au, however (Fig.~\ref{fig:FF-vc_cptsT2}), although the overall step
size is less, the AE approximation captures neither
the step in $v\c^T$ nor in $v\c^W$. The AE again does a reasonable job of
capturing $v\c^{\rm hole}$ although not getting all its structure
correct, similar to the case of the local Rabi excitation in Sec.~\ref{sec:sftcHe-Rabi}. 

\begin{figure}[h]
\begin{center}
\includegraphics[width=0.45 \textwidth]{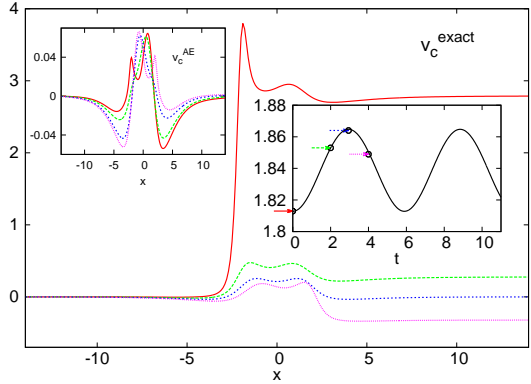}
\end{center}
\caption{The dynamical step shown at 4 times indicated in the inset, and the dominant NOON (inset) in the field-free evolution of Eq.~(\ref{eq:IS}) in 1D He.}
\label{fig:FF_tdnos}
\end{figure}

\begin{figure}[h]
\begin{center}
\includegraphics[width=0.45 \textwidth,height=7cm]{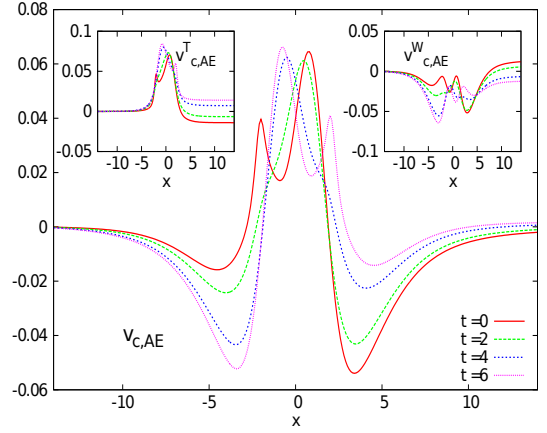}
\end{center}
\caption{The AE correlation potential, $v\c^{\rm AE}$, and its components $v\c^{W,{\rm AE}}$ and $v\c^{T,{\rm AE}}$, in field-free evolution example. The largest adiabatic NOON is 1.94251, 1.94245, 1.94253, 1.94252 at times 0,2,4,6, respectively. }
\label{fig:FF_tdnos_adiacpts}
\end{figure}

We now turn to the question of the NOONs plotted in
Figure~\ref{fig:FF_tdnos} over one period of oscillation. Notice that
initially, the largest occupation numbers are 1.813 and 0.184, which
are not very far from the SSD values of 2 and 0. Despite not deviating
too far from a single-Slater determinant (i.e. being weakly correlated), the step in $v\c$ is really
quite large on the scale of the entire potential, suggesting, as in
the previous section, that the system does not need to wander far from
an SSD for the dynamical step to be important, in contrast to the potential steps found in the ground-state case. We note once again,
that the maximum value of the step appears to appear at local minima
of the NOONs (and vice-versa). 
Finally, Figure~\ref{fig:FF_tdnos_adiacpts} focusses on $v\c^{\rm AE}$ and its components, and shows
that the steps in $v\c^{W,{\rm AE}}$ and $v\c^{T,{\rm AE}}$ oscillate, although on a much smaller scale than the step in the exact dynamical potentials, and moreoever they
largely cancel when added together as noted earlier. The AE NOONs vary very little, and the AE system stays very weakly correlated throughout.

\subsection{1D H2: Resonant energy transfer dynamics}
\label{sec:RET}
\begin{figure}[h]
\begin{center}
\includegraphics[width=0.4\textwidth]{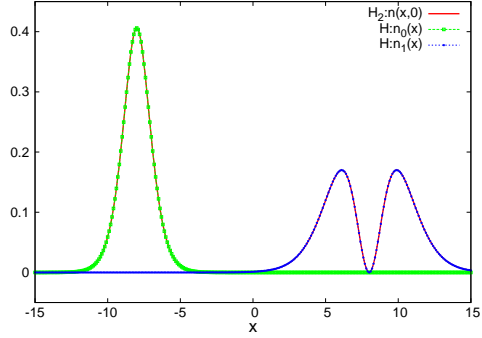}
\end{center}
\caption{Initial density in the 1D H$_2$ molecule $n(x,0)$ (red solid line), compared with 
the ground-state density of a hydrogen atom on the left $n_0(x)$ (green points) and the excited state density of a hydrogen atom on the right $n_1(x)$ (blue points).}
\label{fig:initdens_RET}
\end{figure}

We now consider a case where an excitation transfers over a long
distance but without charge transfer. We place our two soft-Coulomb interacting electrons in a 1D model of the H$_2$
molecule: \ben v\ext(x) = -1/\sqrt{(x-R/2)^2+1} - 1/\sqrt{(x+R/2)^2+1}
\een and take $R=16$ a.u.  The exact ground-state of this molecule has a
Heitler-London nature in the limit of large separation, 
\ben
\Psi^{\rm g.s.}(x,x') = \left(\phi_L(x)\phi_R(x')+\phi_R(x)\phi_L(x')\right)/\sqrt{2}
\een
while the
lowest two singlet excitations become:
\bea
\Psi^{\rm (1)}(x,x') &=& \left(\phi_L(x)\phi^{*}_R(x') + \phi^{*}_L(x)\phi_R(x')+(x\leftrightarrow x'))\right)/2\\
\nonumber
\Psi^{\rm (2)}(x,x') &=&\left(\phi_L(x)\phi^{*}_R(x') - \phi^{*}_L(x)\phi_R(x')+(x\leftrightarrow x'))\right)/2
\eea
where $\phi_{L,R}$ denote the ground-state hydrogen
orbitals on the left and right atoms, and $\phi^{*}_{L,R}$ denote
the excited state atomic orbitals.  The charge-transfer resonances,
H$^+$H$^-$ + H$^-$H$^+$ (in the large separation limit), are found at
higher energies in this model.  We begin with an initial excitation
localized in the right-hand-well, which is specifically a 50:50
combination of the first two excited states, $\Psi(0) =
\left(\Psi^{\rm (1)} +\Psi^{\rm (2)}\right)/\sqrt{2}$. The density is
essentially that of a local excitation on the right atom and the
ground-state on the left and is compared with the hydrogen atom ground
and first excited state densities on each atom in
Figure~\ref{fig:initdens_RET}. The electrons are then allowed to evolve, as in the previous section, with no external field applied. 

\begin{figure}[h]
\begin{center}
\includegraphics[width=0.45 \textwidth]{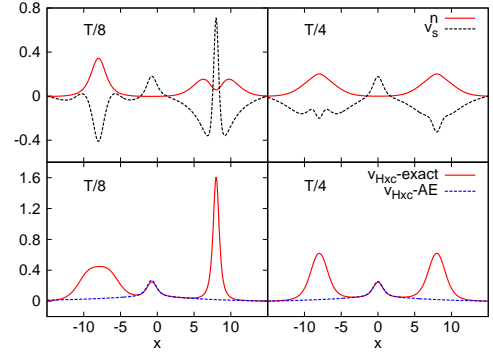}
\end{center}
\caption{Top panels: exact KS potential (black dashed) and density (red solid) at times shown during the resonant energy transfer in the H$_2$ molecule. Lower panels: The exact Hartree-xc potential, $v\H + v\xc$ (red solid) and its AE approximation (blue dashed).}
\label{fig:RET-n_vs_vhxc}
\end{figure}

As the right-hand well de-excites, the density in the left-hand-well
gets excited; the excitation transfers back and forth while the
density remains integrated to one electron on each well at all times. The
density and full KS potential are plotted in
Figure~\ref{fig:RET-n_vs_vhxc} at two times during the energy transfer; $T$ is the period of the dynamics, $T = 2\pi/(E^{(2)} -E^{(1)}) = 5374.84$a.u. After $T/2$ the excitation has transferred completely to the other atom and the pictures at times between $T/4$ and $T/2$ are the same as those between $0$ and $T/4$ but flipped around the $x$-axis. 

Any dynamical step is too small to be observed.
The system seems to be essentially two
one-electron systems in each well, each getting excited then
de-excited; so one might expect that Hartree-xc effects are minimal,
at least locally in each well and that the KS potential would revert
to the external potential in the one-electron regions around each
well. (Certainly, for a time-dependent truly one-electron system, $v\s
= v\ext, v\c=0, v\x = -v\H$). Turning to the lower panels in Figure~\ref{fig:RET-n_vs_vhxc} we
see this is not in fact the case for the exact $v\H + v\xc$.  The AE
$v\H +v\xc$ does show the above described behavior, i.e. it becomes flat in the region in each well in the large separation limit and only the intermolecular midpoint peak remains.  This midpoint peak is similar to
the peak in the ground-state potential in H$_2$ that appears as the
ground-state molecule dissociates~\cite{GB97,TMM09} 
and is a feature
of the kinetic component to the correlation potential, $v\c^T$ (see
shortly). However the {\it exact} $v\H +v\xc$ is certainly nowhere near
becoming flat locally around each well! The interacting system cannot
be thought of as solving a one-electron Schr\"odinger equation in each well: although
locally the density is a one-electron density, the system cannot be described by one orbital in each well. 

\begin{figure}[h]
\begin{center}
\includegraphics[width=0.4\textwidth]{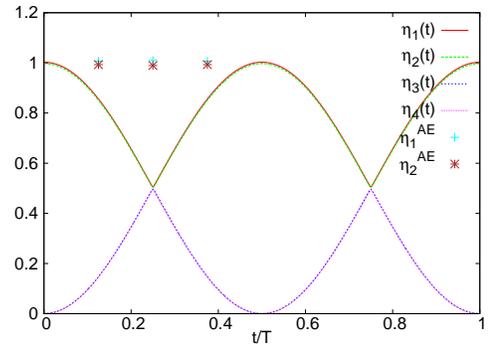}
\end{center}
\caption{The four significant NOONs over a period of oscillation of the energy transfer. The largest two AE NOONs are also shown at discrete times as points.}
\label{fig:RET-tdnos}
\end{figure}

\begin{figure}[h]
\begin{center}
\includegraphics[width=0.45\textwidth]{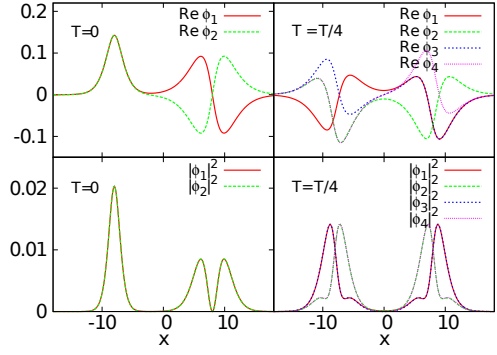}
\end{center}
\caption{(color online) Upper panel: Real part of occupied NOs at t=0 and T/4. NOs appear pairwise and have the structure  $f_1(x) \pm f_2(x)$  as discussed in the text.
Lower panel: orbital densities at same snapshots.}
\label{fig:RET_NO}
\end{figure}

To see this more
precisely, take a look at the NOONs
plotted in Figure~\ref{fig:RET-tdnos} and the NOs themselves, plotted in Fig.~\ref{fig:RET_NO}. 
At the initial time and every half-period, there are two NOs that are equally occupied: in fact these have a bonding and antibonding structure across the molecule, and are identical up to a sign locally in each well, as can be seen from the top left panel of Fig~\ref{fig:RET_NO}. At these times therefore one orbital describes the dynamics in each well, and the problem resembles the stretched H$_2$ molecule  (Heitler-London). In fact at $t=0,T/2$ the exact $v\H + v\xc$ does become flat locally in the region of each atom (not shown). 
Away from the initial time and
half-periods, more than two natural orbitals are significantly occupied. 
At a quarter-period, when there is equal excitation on both wells,
four natural orbitals are equally occupied and these are shown on the right panels of Fig.~\ref{fig:RET_NO}.
Around each well, two of
the four largest natural orbitals have essentially identical
densities; pairwise, they have the structure of $f_1(x) \pm f_2(x)$ where
$f_{(1,2)}(x)$ is a function localized on the left(right), but, importantly,
different pairs have different $f_i(x)$. This means that the electron
localized in one well is being described by four orbitals, which are
pairwise essentially identical, but quite distinct from the other
pair. That is, each electron is locally described by two distinct
functions with comparable weights: definitely not a one-electron
dynamics, despite being a one-electron density. As a result the exact
$v\H+ v\xc$ does not vanish locally around each well as would be the
case for one-electron systems (time-dependent or ground-state), see Figure~\ref{fig:RET_NO}.
The excitation--de-excitation process in each well cannot be described by a pure state ($a_1\varphi_1 +a_2\varphi_2$).
 Note that the AE NOONs stay constant and extremely
close to 1. The AE NOs (not shown) also have the symmetric/antisymmetric combination
structure $g_1(x) \pm g_2(x)$, but around each well the two orbitals are
essentially identical, like for the exact case at the initial time. Each electron in the AE approximation is
therefore described by one function around each well, and so the system does behave locally as a one-electron system, and hence in the AE approximation the Hartree-xc potential vanishes locally around each well.

\begin{figure}[h]
\begin{center}
\includegraphics[width=0.45 \textwidth]{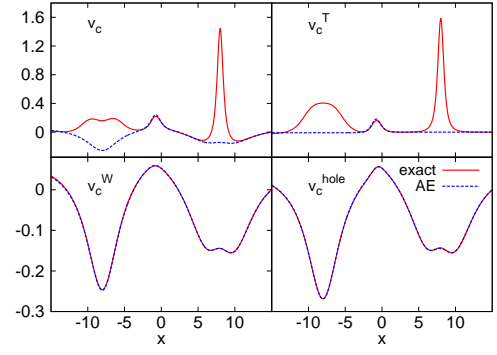}
\end{center}
\caption{(color online) Components of $v\c$ at  $t=T/8$ for the resonant energy transfer model.}
\label{fig:RET_T1o8}
\end{figure}

\begin{figure}[h]
\begin{center}
\includegraphics[width=0.45\textwidth]{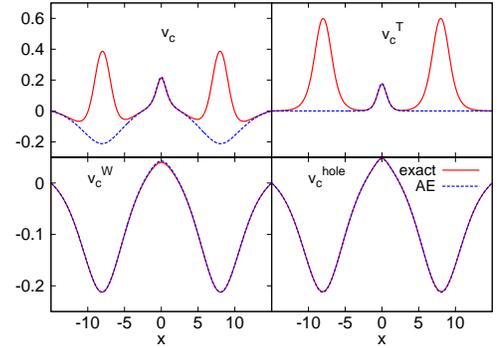}
\end{center}
\caption{(color online) Components of $v\c$ at $t=T/4$ for the resonant energy transfer model. }
\label{fig:RET_T1o4}
\end{figure}

Figures~\ref{fig:RET_T1o8}--\ref{fig:RET_T1o4} plot the correlation
potential and its components $v\c^T, v\c^W, v\c^{\rm hole}$ at two
times during the energy transfer. We observe that the AE approximation
is consistently essentially exact for the interaction contributions
$v\c^W$, and $v\c^{\rm hole}$, which in fact exactly cancel the
Hartree-exchange potential locally: $v\c^{W,{\rm AE}} = v\c^W$ and
$v\H +v\x + v\c^{W} = 0$ locally in each well. We can understand this,
since being a one-electron density in each well, there should be no
self-interaction from the Coulomb interaction, so the interaction
contribution $v\c^W$ must just cancel the Hartree and exchange
potential. (Globally we have a
two-electron system so $v\x = -v\H/2$ instead of completely
cancelling Hartree; $v\c^{W}$ then steps in to complete the job, which
is called a static correlation effect and also occurs in the
ground-state of stretched molecules~\cite{GB97,TMM09}).  The entire
non-trivial structure of $v\c$ is in its kinetic component $v\c^T$,
and is due to the effect discussed in the last paragraph, and is
completely missed by the AE approximation, $v\c^{T,{\rm AE}} =0$
locally in each well. Similar behavior appears at other times  that are not
shown.

\section{Conclusions and Outlook}
\label{sec:conclusions}
We have presented a decomposition of the exact time-dependent xc
potential into kinetic and interaction components, similar to the
corresponding decomposition in the ground-state which has proven
useful for understanding features of the ground-state xc
potential~\cite{BBS89,GLB94,GLB96,GB96}.  We have made the first
studies of these components for three different non-perturbative dynamical situations
and compared them to their adiabatically-exact counterparts: resonant
Rabi oscillations in a 1D He model, field-free dynamics of a
superposition state in the 1D He atom, and resonant excitation energy
transfer in a 1D H$_2$ molecule. We found that the step and peak
structures in the correlation potential that were recently found in a
range of dynamical situations~\cite{EFRM12, FERM13} are largely, but
not exclusively, contained in the the kinetic component $v\c^T$ of the
correlation potential. Even in the absence of step structure, $v\c^T$
is typically considerably worse approximated by the
adiabatically-exact approximation compared to the other terms in
situations far from the ground-state. The case of resonant energy
transfer in the 1D H$_2$ molecule was an extreme case where one
electron lives in each atomic well, but the excitation transferring
back and forth between the atoms via the Coulomb interaction led to a
large non-adiabatic component of $v\c^T$ in each well, a signature of the
fact the the dynamics in each well cannot be described by a single
orbital. In this case the AE approximation for $v\c^W$ was practically exact. 

Step structures in the ground-state are associated with strong
deviation from a SSD, but we found that the relationship between the time-dependent
NOONs and the dynamical step is not so simple. There may be strong
static correlation in the system, while there is no step, and the step
may be large even when the system is weakly correlated. Instead, we found that the oscillations of the
dynamical step size are associated with oscillations in the time-dependent
NOONs, interestingly, and further explorations of the trends and dependences in
different cases will be carried out.

The examples studied suggest that one may get away with an adiabatic
approximation for $v\c^{\rm hole}$, while the error from an adiabatic
approximation to $v\c^W$ and particularly $v\c^T$ would be much
larger. Still, the importance of each of these terms in influencing
the dynamics has yet to be studied. A point of future study would be
to self-consistently propagate separately under the three components
mentioned to gauge their relative importance on the resulting
dynamics.  The comparison with the dipole dynamics given by adiabatic
approximations (Fig.~\ref{fig:dipoles} for the Rabi oscillation in 1D
He model, Fig. 6 of Ref.~\cite{EFRM12} for the field-free evolution in
the same potential) certainly suggests that the non-adiabatic effects
are important. To disentangle the effect of the adiabatic
approximation itself and the choice of the ground-state approximation,
a self-consistent propagation under the AE approximation would be
enlightening, and is an important avenue for future work.

The equation for the exact xc potential, Eq.~(\ref{eq:vxc}) is valid for
$N$-electrons, and points directly to what approximations to the xc
potential are attempting to model: gradients of the correlated part of
the one-body density matrix (kinetic component), and Coulomb-type
integrals of the xc hole (interaction components). The equation gives the xc
field, i.e. the gradient of the xc potential, so even a local model of
the right-hand-side could give  a spatially non-local potential. The notion
is somewhat reminiscent of the motivations of time-dependent
current-density functional theory~\cite{VK96}. The step structure in the potential requires a non-local density-dependence, but the
electric field it represents, $\nabla v\c$, is quite localized.

Due to the one-body nature of the KS evolution operator, the form of
the KS state at any time remains the same as that chosen for the KS
initial state; orbitals composing the KS initial state evolve in time
with no change in their occupations.  Throughout this work, we have
taken the KS initial state to be a SSD consisting of one
doubly-occupied orbital.  In principle, more general initial KS states
may be chosen provided they have the same density and the first
time-derivative of the density as that of the interacting system. The
question then arises as to whether, if stuck with an adiabatic
approximation, is there a choice of KS initial state that the
adiabatic approximation works best for?  In fact, a judicious choice
of the initial KS state can lessen the error that an adiabatic
approximation can make~\cite{EM12,RNL13}. For example, Ref.~\cite{EM12}
considered the interacting system beginning in the first excited
singlet state of the 1DHe atom: there the adiabatic approximation to
$v\xc$ initially gives a far closer approximation to the exact xc
potential if the KS state is chosen as a double-Slater-determinant
with one ground-state orbital and the other a lowest excited-state
orbital, instead of the usual choice of a doubly-occupied
orbital. This suggests to choose a KS initial state with a
configuration similar to that of the true initial state to minimize
the error of an adiabatic approximation at least at short times.  On
the other hand, when the interacting system starts in its initial
ground-state, then the adiabatic approximation has least error
initially if the KS initial state is also chosen as a KS ground-state:
with such a choice, the adiabatically-exact approximation is exact at
first. However as time evolves, the interacting state may change its
form dramatically, e.g. in the example A shown in the present paper,
the interacting state starts off in its ground-state,
weakly-correlated, but evolves over time to an excited singlet-state
that minimally requires a two-determinant description. In this case,
beginning with a KS SSD, as we have done, is the best choice for the
adiabatic approximation at short times, however as the excited state
is reached, it becomes increasingly poor. A question for future
research is whether, for a given known structure of the evolution of
the interacting state, there is an optimal choice for the form of the
KS wavefunction such that errors in adiabatic approximation are
minimized throughout the evolution.

How significant are the structures found in the non-adiabatic parts of
$v\c^T$ and $v\c^W$, and their impact on the ensuing dynamics, for
realistic three-dimensional systems of more than two electrons remains
to be tested; this is clearly a more challenging numerical task. The
analysis in terms of the kinetic and interaction contributions of the xc
potential should prove useful to deepen our understanding of
time-dependent electron correlation, and eventually to modelling
non-adiabatic effects accurately.

{\it Acknowledgments} Financial support from the National Science
Foundation CHE-1152784 (for KL), Department of Energy, Office of
Basic Energy Sciences, Division of Chemical Sciences, Geosciences and
Biosciences under Award DE-SC0008623 (NTM., JIF), the European
Communities FP7 through the CRONOS project Grant No. 280879 (PE),
a grant of computer time from the CUNY High Performance Computing
Center under NSF Grants CNS-0855217 and CNS-0958379, and the RISE program at Hunter College, Grant GM060665 (ES),  are gratefully
acknowledged.


\begin{thebibliography}{99}
\bibitem{RG84}
E. Runge and E.K.U. Gross, Phys. Rev. Lett. {\bf 52}, 997 (1984).

\bibitem{TDDFTbook}
{\it Fundamentals of Time-Dependent Density Functional Theory, (Lecture Notes in Physics 837)},   eds. M.A.L. Marques, N.T. Maitra, F. Nogueira, E.K.U. Gross, and A. Rubio, (Springer-Verlag, Berlin, Heidelberg, 2012); and references therein.

\bibitem{Carstenbook}
{\it Time-Dependent Density-Functional Theory: Concepts and Applications}, C. A. Ullrich, (Oxford, 2012).



\bibitem{SSYI12}
Y. Shinohara et al., J. Chem. Phys. {\bf 137}, 22A527 (2012). 

\bibitem{Rozzi13}
C. A. Rozzi et al., Nature Commun. {\bf 4} 1602 (2013)

\bibitem{PDP09}
O. V. Prezhdo, W. R. Duncan, V. V. Prezhdo,
Prog. Surf. Sci. {\bf 84}, 30 (2009). 

\bibitem{BKPB11}
I. Bocharova et al., Phys. Rev. Lett. {\bf 107}, 063201 (2011). 

\bibitem{BV12}
S. Bubin and K. Varga, Phys. Rev. B. {\bf 85}, 205441 (2012). 

\bibitem{C13}
A. Castro, ChemPhysChem. {\bf 14}, 1488 (2013). 

\bibitem{TGK08}
M. Thiele, E.K.U. Gross, S. K\"ummel, Phys. Rev. Lett. {\bf 100}, 153004 (2008)

\bibitem{EFRM12}
P. Elliott, J. I. Fuks, A. Rubio, and N. T. Maitra, Phys. Rev. Lett. {\bf 109}, 266404 (2012).

\bibitem{RG12}
J. D. Ramsden and R. W. Godby, Phys. Rev. Lett. {\bf 109}, 036402 (2012).


\bibitem{FERM13}
J. I. Fuks, P. Elliott, A. Rubio, and N. T. Maitra, J. Phys. Chem. Lett. {\bf 4}, 735 (2013).

\bibitem{RNL13}
M. Ruggenthaler, S. E. B. Nielsen, and R. van Leeuwen, Phys. Rev. A {\bf 88}, 022512 (2013) 

\bibitem{BBS89}
M. A. Buijse, E. J. Baerends, J. G. Snijders, Phys. Rev. A. {\bf 40}, 4190 (1989). 

\bibitem{GLB94}
O. Gritsenko, R. van Leeuwen, and E.J. Baerends, J. Chem. Phys. {\bf 101}, 8955 (1994)

\bibitem{GLB96}
O. Gritsenko, R. van Leeuwen, and E. J.  Baerends, J. Chem. Phys. {\bf 104}, 8535 (1996)

\bibitem{GB96}
O. Gritsenko and E. J. Baerends, Phys. Rev. A. {\bf 54}, 1957 (1996). 

\bibitem{RB09b}
M. Ruggenthaler and D. Bauer, Phys. Rev. A {\bf 80},052502 (2009) 

\bibitem{L99}
R. van Leeuwen, Phys. Rev. Lett. {\bf 82}, 3863 (1999). 


\bibitem{MB01}
N. T. Maitra and K. Burke, Phys. Rev. A. {\bf 63}, 042501 (2001).


\bibitem{EM12}
P. Elliott, N. T. Maitra, Phys. Rev. A {\bf 85}, 052510 (2012).

\bibitem{TMM09}
D. J. Tempel, T. J. Martinez, N. T. Maitra, J. Chem. Theory and Comput. {\bf 5}, 700 (2009).



\bibitem{LEM13}
K. Luo, P. Elliott, N. T. Maitra, Phys. Rev. A. {\bf 88}, 042508 (2013).

\bibitem{RNL12}
M. Ruggenthaler, S. Nielsen, R. van Leeuwen, 
Europhys. Lett. {\bf 101}, 33001 (2013).


\bibitem{octopus}
Castro, A. {\it et~al.}. Octopus: A Tool for the Application of Time-Dependent Density
Functional Theory. {\it Phys. Stat. Sol. (b)} {\bf 2006}, {\it 243}, 2465--2488.

\bibitem{octopus2}
Marques, M. A. L.; Castro, A.; Bertsch, G. F.; Rubio, A. Octopus: A First-Principles Tool
for Excited Electron-Ion Dynamics. {\it Comp. Phys. Comm.} {\bf 2003}, {\it 151}, 60--78.

\bibitem{FHTR11}
Fuks, J. I.; Helbig, N.; Tokatly, I.V.; Rubio, A. 
{\it Phys. Rev. B.} {\bf 2011}, {\it 84}, 075107. 

\bibitem{RB09}
M. Ruggenthaler and D. Bauer, Phys. Rev. Lett. {\bf 102}, 233001 (2009).


\bibitem{FM14a}
 J. I. Fuks and N. T. Maitra, submitted to Phys. Chem. Chem. Phys. (2014).

\bibitem{RN11}
S. Raghunathan and M. Nest, J. Chem. Theory Comput. {\bf 7}, 2492 (2011). 

\bibitem{GB97}
O. V. Gritsenko and E. J. Baerends, Theor. Chem. Acc. {\bf 96}, 44 (1997). 


\bibitem{VK96}
G. Vignale and W. Kohn, Phys. Rev. Lett. {\bf 77}, 2037 (1996).

\end{thebibliography}
\end{document}